\documentclass[a4paper,10pt]{article}
\usepackage[utf8]{inputenc}

\usepackage{amsmath}
\usepackage{amssymb}
\usepackage{hyperref}
\usepackage{graphicx}
\usepackage{verbatim}
\usepackage{geometry}
\usepackage{tikz}
\tikzstyle{every picture}=[level distance = 8mm, baseline=-0.5ex]
\tikzstyle{prop}=[shape=circle,minimum size=6mm, draw=black!80, fill=green!30]

\renewcommand{\d}{\text{d}}

\newcommand{\C}{\mathbb{C}}

\newcommand{\Z}{\mathbb{Z}}

\begin{document}

\title{Alien calculus and a Schwinger--Dyson equation: two-point function with a nonperturbative mass scale}
\author{Marc~P.~Bellon${}^{1,2}$, Pierre~J.~Clavier${}^{3}$\\
\normalsize \it ${}^1$ Sorbonne Universités, UPMC Univ Paris 06, UMR 7589, LPTHE, 75005, Paris, France\\
\normalsize \it $^2$ CNRS, UMR 7589, LPTHE, 75005, Paris, France\\
\normalsize \it $^3$ Potsdam Universit\"at, Mathematik, Golm, Deutschland }

\date{}

\maketitle

\begin{abstract}
Starting from the Schwinger--Dyson equation and the renormalization group equation for the massless Wess--Zumino model,
we compute the dominant nonperturbative contributions to the anomalous dimension of the theory, which are related by alien calculus to
singularities of the Borel transform on integer points. 
The sum of these dominant contributions has an analytic expression.
When applied to the two-point function, this analysis gives a tame evolution in the deep euclidean domain at this approximation level,
making doubtful the arguments on the triviality of the quantum field theory with positive \(\beta\)-function.  On the other side, we have a singularity of the 
propagator for timelike momenta of the order of the renormalization group invariant scale of the theory, which has a nonperturbative relationship with the 
renormalization point of the theory.  
All these results do not seem to have an interpretation in terms of semiclassical analysis of a Feynman path integral.
\end{abstract}

\textbf{Mathematics Subjects Classification:} 81Q40, 81T16, 40G10.


\section*{Introduction}
Quantum field theory required the finest element of mathematical physics for its development, pushing at time the development of new aspects of mathematics.  
The analyticity requirements have been instrumental in the theory of the holomorphic functions of many variables and their analyticity domains, 
with such high points as the edge of the wedge theorem of Bogolubov or the Bros--Iagolnitzer transform used for the definition of the analytic 
wave front. Type III factors are everywhere and the instrument of their classification stems from the KMS condition, introduced for the study of 
finite temperature theory, which was found to coincide with the Tomonoga condition. Perturbative expansions and their renormalization require 
subtle combinatorics and the introduction of Hopf algebras allows one to clarify and make more explicit the classical proofs of renormalizability. 
Using the formalism of groupoids may be useful to reduce the burden of controlling the effect of the symmetry factors. Evaluating Feynman 
integrals requires numbers which can be periods, with the action of a motivic Galois group and links with many conjectures in algebraic geometry.
Constructive theory has been able to show that some of these theories can be given a precise mathematical sense, but has failed to address 
the most relevant ones for our understanding of the physical world, the four dimensional gauge theories. 

In fact, due to the specificity of renormalization, the perturbative series is the richest source of information on quantum field theory.
Other approaches deal necessarily with a regularized version of the theory which lacks many of the structural properties of the final theory.  A precise 
definition will therefore seek to define the relevant Green functions from their perturbative series, but the procedure cannot be straightforward, 
since it has been long recognized that quantum field theories (QFT for short) cannot depend holomorphically on the coupling parameter in a full 
neighborhood of the origin and therefore the perturbative series is at most asymptotic. A number of works have tended to show that the growth of 
the terms of the formal perturbative series is slow enough to allow the definition of a Borel transform around the origin.  This is however 
only the first step in the definition of a sum for the perturbative series.  One must also be able to analytically continue this Borel 
transform up to infinity and furthermore verify that near infinity, this analytically continued Borel transform does not grow 
faster than any exponential function.

Defining an analytic 
continuation seems a formidable task, but in the cases where the functions obey equations, analog equations for their Borel transform can be 
deduced and, with the help of alien calculus, used to constrain the possible singularities of the Borel transform.  Singularities may appear on 
the positive real axis, preventing a straightforward application of the Laplace transform.  One could resort to lateral summation, using a 
shifted integration axis, but this means that the reality properties of the solution are lost.  This can be an advantage in some situations, 
where the imaginary part of a perturbed energy signals the possible decay of a metastable state to the continuum, but unitarity could be at risk.  However a suitable average of the analytic continuations circumventing the singularities on the real axis can be used to define a sum which both is real and respects the equations.

In this work, we will therefore argue that quantum field theory cannot dispense with the whole body of work on summation methods which has been 
thoroughly expanded by Jean Ecalle~\cite{Ecalle81,Ecalle81b,Ecalle81c}.  As in previous works~\cite{BeCl14}, we will base our considerations on computations in a specific model, 
but we think that they reveal phenomena at work in most exactly renormalizable field theories.

Although the insight about these summation methods comes from the study of Borel transforms which form an algebra under convolutive products, 
much can be done while remaining in the domain of more or less formal series.   In particular, formal transseries solutions allow to 
recognize the possible forms of alien derivatives.  It is then possible to express the solution in terms of transmonomials, special functions 
with simple alien derivatives, so that we never have to explicitly refer to the Borel transforms in computations.  The Borel-transformed functions are however 
what justify the different computations.  

This article is divided as follows: first we give a short introduction to some concepts of resurgence theory that are of importance for our work. 
Then some previous results of \cite{BeSc08,BeCl13,BeCl14} that are to be developed further are recalled. We then compute the leading terms 
for every exponentially small terms in the anomalous dimension in the subsequent section. They are shown to sum up to a known analytic function of the 
nonperturbatively small  quantity \(e^{-r}\).  Alien calculus is then used to deduce from the computed terms the singularities of the Borel transform, while 
the first singularity of the Borel transform is used to constrain a free parameter in the previous paragraphs.
Finally, we apply the same process to the two-point function of the theory and see that the nonperturbative terms can get multiplied by powers of the momentum. The resulting function has a singularity for a timelike momentum, which fixes a nonperturbative mass scale, while the euclidean side is completely tame.  The way such a result could appear in the process of Borel summation gives further indications on the analyticity domain of the theory.

\section{Elements of resurgence theory}

\subsection{Borel transform and Borel resummation}

A very nice introduction to the subject of Borel transform and resurgence theory is \cite{Sa14}. Here we will follow the presentation of \cite{Bo11}, 
with some additional material needed for our subject.

A simple definition of the Borel transform is to say that it is a morphism between two rings of formal series, defined as being linear and its value on monomials:
\begin{eqnarray}
             \mathcal{B}: a\mathbb{C}[[a]] & \longrightarrow & \mathbb{C}[[\xi]] \\
\tilde{f}(a) = a\sum_{n=0}^{+\infty}c_na^n & \longrightarrow & \hat{f}(\xi) = \sum_{n=0}^{+\infty}\frac{c_n}{n!}\xi^n. \nonumber
\end{eqnarray}
This definition is extended to noninteger powers by
\begin{equation*}
 \mathcal{B}\left(a^{\alpha+1}\right)(\xi) := \frac{\xi^{\alpha}}{\Gamma(\alpha+1)}.
\end{equation*}
In the case where $\alpha$ is an integer, we find back the above definition.

Then the point-like product of formal series becomes the convolution product of formal series, or of functions when the Borel transform is the 
germ of an analytic function. It is therefore consistent to define the Borel transform of a constant as the formal unity of the convolution product (i.e.,
the Dirac ``function''):
\begin{equation}
 \mathcal{B}(1) := \delta.
\end{equation}

In the following, we will restrict ourselves to the cases where $\hat f$ is the germ of an analytic function at the origin, endlessly continuable on 
$\mathbb{C}$, i.e., that on any line $L$ of $\mathbb{C}$ a representative of the germ $\hat f$ has a countable number of singularities and is continuable along 
any path obtained by following $L$ and avoiding the singularities by going over or below them. Moreover, we will assume that these singularities are 
algebraic, that is to say that they are not more singular than a pole. More precisely, if $\xi_0$ is a singularity of $\hat\phi$, we shall have
\begin{equation}
 \exists\alpha\in\mathbb{R}\mid  |(\xi-\xi_0)^{-\alpha}\hat\phi_r(\xi)|\underset{\xi\rightarrow\xi_0}{\longrightarrow}0.
\end{equation}
The supremum of the $\alpha$ for which the above holds will be called the order of $\hat\phi$ at $\xi_0$. Notice that if $\alpha$ is positive, 
it is important that the condition is not directly on \(\hat\phi\) but on \(\hat\phi_r\), obtained from \(\hat\phi\) by subtracting a suitable polynomial in~\(\xi\).  {A typical example for \(\alpha\) a positive integer is \(\hat\phi_r(\xi) \simeq (\xi-\xi_0)^\alpha \ln(\xi-\xi_0) \), but there are no reason for \(\hat\phi\) itself to have a zero at \(\xi_0\).  Alternatively, the singularity can be characterized by the behavior of the difference of the function and its analytic continuation after looping around \(\xi_0\), but this does not work for poles.}
It is even possible to have singularities with an infinite order, but in our applications, the order will always be a finite rational number.

Let us call $\widehat{RES}$ the space of germs of analytic functions at the origin endlessly continuable on $\C$ and $\widetilde{RES}\subset a\mathbb{C}[[a]]$ the set of formal series whose image under the Borel transform is in
$\widehat{RES}$. When working with elements of $\widehat{RES}$ we will say that we are in the convolutive model, while $\widetilde{RES}$ will be called the 
formal model. We will also say that we are in the Borel plane and the physical plane, respectively.

There exists an inverse to the Borel transform: the Laplace transform.  {For $\phi\in\widehat{RES}$ we write $\hat\phi$ for the analytic continuation of the Borel transform of $\phi$. The definition of the Laplace transform 
on $\hat\phi$} involves a certain direction $\theta$ in the complex plane:
\begin{equation}
 \mathcal{L}_{\theta}[\hat\phi] := \int_0^{e^{i\theta}\infty}\hat\phi(\zeta)e^{-\zeta/a}\d\zeta.
\end{equation}
It is well defined for at least some values of \(a\) if \(\hat\phi\) is smaller than some exponential in the direction \(\theta\). The resummation operator in the direction $\theta$ is the composition of the Borel transform and the Laplace transform in the direction \(\theta\):
\begin{equation}
 S_{\theta} := \mathcal{L}_{\theta}\circ\mathcal{B}.
\end{equation}
If $\hat\phi$ does not have any singularities in the directions \(\theta\) between $\theta'$ and $\theta''$ included and satisfies suitable exponential bounds at infinity in this sector, different $\mathcal{L}^{\theta}[\hat\phi]$ coincide wherever they are both defined through Cauchy's theorem, so that they define a single analytic function on the sector delimited 
by $\theta' -\pi/2$ and $\theta''+ \pi/2$, which is a possible resummation of the formal series \(\tilde \phi\). We see that the different resummations are defined in sectors whose limits depend on the singularities of the Borel transform, but their definition domains have nontrivial intersections. 

In the following, we will need more general objects than the elements of \(\widehat{RES}\) and the corresponding $\widetilde{RES}$. 
We define simple resurgent symbols with an additional variable \(\omega \in \mathbb{C}\) with the meaning that the corresponding object in the formal or geometric models get multiplied by \(e^{-\omega/a}\), so that:  
\begin{equation*}
 \dot\phi^{\omega} := e^{-\omega/a}\phi\in\dot{\widetilde{RES}} \supset \widetilde{RES}
\end{equation*}
In the formal model, linear combinations of simple resurgent symbols are simple examples of transseries. One can define more general transseries, but this would lead us away from our topic.  {A very pleasant introduction to transseries, very accessible to physicists is} \cite{Ed09}.

Finally, the operator $S_{\theta}$ extends to \(\dot{\widehat{RES}}\) through the formula
\begin{equation}\label{SthetaDot}
 S_{\theta}[\hat\phi^{\omega}](a) := e^{-\omega/a}\mathcal{L}_{\theta}[\hat\phi](a)
\end{equation}
and by linear extension.

\subsection{Stokes automorphism and Alien derivative}

Now let $\phi\in\widetilde{RES}$ be such that $\hat\phi\in\widehat{RES}$ has singularities in the direction $\theta$. Then we define the lateral 
resummations $S_{\theta\pm}$ as the usual one but with the Laplace transform \(\mathcal{L}_{\theta\pm}\) involving integrals avoiding the singularities by going above (for 
$\mathcal{L}_{\theta+}$) or below (for $\mathcal{L}_{\theta-}$) all of them. They correspond to the limit of \(S_{\theta'}\) when \(\theta'\) tends to \(\theta\) either from above or from below.\footnote{{
In most applications, the possible arguments of the positions of the singularities form a discrete set, so that the different \(S_{\theta'}\) define the same analytic function for an open set of values of \(\theta'\) and we do not really need to take the limit in the definition of \(S_{\theta\pm}\).  However, it is possible to have singularities for example at the positions of all Gaussian integers \(\mathbb{Z} + i \mathbb{Z}\), in which case the limiting procedure is unavoidable.}}
\begin{equation}
 S_{\theta\pm}[\phi](a) := \int_0^{e^{i(\theta\pm\varepsilon)}\infty}\hat\phi(\zeta)e^{-\zeta/a}\d\zeta.
\end{equation}
The extension of the lateral resummations to the elements of $\dot{\widetilde{RES}}$ is similar to the extension of the regular ones given by equation~(\ref{SthetaDot}).

Now, the key point is that the lateral resummation are linked by the so-called Stokes automorphism in the direction $\theta$, written 
$\mathfrak{G}_{\theta}$.
\begin{equation} \label{def_Stokes}
 \mathcal{L}_{\theta+} \circ \mathfrak{G}_{\theta} = \mathcal{L}_{\theta-}.
\end{equation}
We clearly have $\mathfrak{G}_{\theta}:\dot{\widehat{RES}}\longrightarrow\dot{\widehat{RES}}$. Since both \(\mathcal{L}_{\theta+}\) and \(\mathcal{L}_{\theta-}\) are algebra morphisms from the convolutive algebra in the Borel plane to the algebra of functions, \(\mathfrak{G}_{\theta}\) is an automorphism of the convolution algebra \(\dot{\widehat{RES}}\). This automorphism encodes how the 
function ``jumps'' when the direction of integration crosses a line of singularities (called a Stokes line), already in the extended convolutive model. It can be decomposed in homogeneous components that shift the exponents of the extended models by the complex numbers \(\omega\) which belong to the direction \(\theta\): they are called the lateral alien operators \(\Delta^+_\omega\). The action of the alien operator \(\Delta_\omega^+\) on \({\hat\phi}^\sigma\) is linked to the singularity of \({\hat\phi}^\sigma\) in \(\omega\) and carries the index \(\omega+\sigma\).  The fact that \(\mathfrak{G}_{\theta}\) is an automorphism translates in the following relations for its components:
\begin{equation} \label{rel_lateral}
	\Delta^+_\omega (\hat f \star \hat g) = \sum_{\omega' + \omega'' = \omega} \Delta^+_{\omega'} f \star \Delta^+_{\omega''} g,
\end{equation}
where the sum includes the cases where \(\omega'\) or \(\omega''\) is 0, and \(\Delta^+_0\) is defined to be the identity.

Since the relation \eqref{rel_lateral} is not very simple, we use the logarithm of the Stokes automorphism, which is a derivation. The homogeneous components of this logarithm are therefore also derivations, which are called alien derivatives.  More precisely
\begin{equation} \label{def_alien}
 \mathfrak{G}_{\theta} = \exp\left(\sum_{\omega\in\Gamma_{\theta}}\Delta_{\omega}\right),
\end{equation}
with $\Gamma_{\theta}$ the singular locus of the function of interest in the direction $\theta$. The alien derivatives and lateral alien 
derivatives are linked by the equivalent relations
\begin{align*}
  & \Delta_{\omega_n}^+ = \sum_{p=1}^n\sum_{\omega_1+\cdots+\omega_p=\omega_n}\frac{1}{p!}\Delta_{\omega_1}\cdots\Delta_{\omega_p} \\
  & \Delta_{\omega_n} = \sum_{p=1}^n\sum_{\omega_1+\cdots+\omega_p=\omega_n}\frac{(-1)^{p-1}}{p}\Delta_{\omega_1}\cdots\Delta_{\omega_p}
\end{align*}
with all the $\omega_i$s on the same half-line from the origin to infinity.

As their names suggest, the alien derivatives are indeed derivations for the convolution product. We shall not give a proof here (we refer the 
reader to \cite{Sa14} for such a proof), but this fact shall not come as a surprise. Indeed, it is well-known that the operator $\mathcal{A}$ 
acting on smooth function as a translation
\begin{equation*}
 \mathcal{A}[f](x) = f(x+1)
\end{equation*}
induces an automorphism on the space of functions. And we can write
\begin{equation*}
 \mathcal{A} = \exp\left(\frac{\d}{\d x}\right)
\end{equation*}
thanks to the Taylor expansion for analytic functions, so that it appears as the exponential of a derivation. Since the Stokes automorphism can be viewed as a translation across a Stokes line, it is understandable that its 
logarithm is a derivative.

The alien operators are a priori defined in the convolutive model, but it is convenient to extend them to $\dot{\widetilde{RES}}$ by
\begin{equation}
 \mathcal{B}[\Delta_{\omega}\phi] := \Delta_{\omega}\hat\phi,
\end{equation}
and similarly for $\Delta_{\omega}^+$. The alien derivatives, so extended to formal series, become derivatives for the point-like product, since the point-like product 
of functions becomes the convolution product in the Borel plane. 

The alien derivatives have other useful properties. The most important one is 
\begin{equation} \label{commutation_alien_usuelle}
 \Delta_{\omega}\partial_z = \partial_z\Delta_{\omega}
\end{equation}
in the formal model, with $z=1/a$. A proof of this result
can be found in \cite{Sa14}, and another in the general case (which is of interest for us) is in \cite{Sa06}. The commutation with the ordinary derivative is not so simple if we consider the alien derivative to act in \(\widehat{RES}\) and not in the dotted model. In fact, in many texts, what we denote simply as \(\Delta_\omega\) is denoted \(\dot\Delta_\omega\).

\subsection{Real resummations}

Stokes operators can be a part of the study of the monodromy around singular points of a differential equation, but it may happen that the Borel transform $\hat\phi$ has singularities in the direction $\theta$ in which we are interested to perform a 
resummation, typically the direction $\theta=0$ for a series with only real coefficients. In this case, the lateral resummations get imaginary parts. This can be a nice feature when the imaginary part of 
the energy corresponds to the decay probability through tunneling of a state, but generally, we would like to obtain a real solution for a physical 
problem.  The issue is that the simple mean of the two different lateral summations is real, but it is nevertheless not satisfying: we would like this 
real resummation to satisfy the same equations as the formal solution and this can only be ensured if the convolution product of the means is the mean of the convolution products. 
The only way to ensure this is to take a suitable combination of the analytic continuations of the function 
along all the paths that can be taken,  going above or below each of the singularities.

It has been known for a long time that such a real solution is given by the so-called median resummation, a fact that have been shown explicitly in 
 \cite{AnSc13}. A possible expression for this median resummation is
\begin{equation}
 S_{\text{med}} := S_{\theta-}\circ\mathfrak{G}_{\theta}^{1/2} = S_{\theta+}\circ\mathfrak{G}_{\theta}^{-1/2}
\end{equation}
where the power of the Stokes automorphism is defined from a natural extension of the definition \eqref{def_alien}:
\begin{equation}
 \mathfrak{G}_{\theta}^{\nu} := \exp\left(\nu\sum_{\omega\in\Gamma_{\theta}}\Delta_{\omega}\right).
\end{equation}
Since the Stokes operator \(\mathfrak{G}_{\theta}\) is an automorphism, its powers are also automorphisms and the median resummation respects products as a composition of operations preserving products.  

Returning to the definition of the Stokes automorphism, it can be seen that \(\Delta^+_\omega\) corresponds to taking the difference between two possible analytic continuation of the Borel transform beyond the point \(\omega\).  The different alien derivatives can also be computed as combinations of different analytic continuations of the Borel transform, going above or below the different singularities (but without ever going backwards).  
The median summation likewise is a suitable average of the different possible analytic continuations of the Borel transform. 
When we go beyond a singularity \(\omega\), we must take a different combination of analytic continuations of the Borel transform: the function we will integrate in the Laplace transform has therefore singularities at the points~\(\omega\) that cannot be avoided and result in nonperturbative contributions to the resummed function.

The square root of the Stokes operator is simple, but since it gives quite an important weight to paths which cross the real axis a large number of times, the obtained average may grow faster than the lateral values.  
In~\cite{Ec92}, Ecalle shows how one could circumvent this problem through accelerations, which allow to reduce the ambiguity between lateral summations from \(1{/}\!\exp(z)\) to \(1{/}\!\exp(\exp(\dots(z)\dots)) \), with theoretically any finite composition of exponentials, through the control of `emanated' resurgence.
However, other averages are possible, the organic averages, still compatible with the convolution product, which are essentially no larger than the lateral determinations and therefore allow us to avoid this whole procedure. 
In any cases, these averages still define from the Borel transform a function which is real on the real axis and has singularities on the real axis so that the sum is only defined in the positive half-plane, since it is impossible to relate this integral to others on different integration axis.

At the approximation level we will reach in the  present work, such subtleties will not have a clear effect.  However, they can become important if we are to improve on our treatment of these nonperturbative contributions.

\section{Rehearsal}
We are still working with the model used in our previous investigations~\cite{BeSc08,BeCl13,BeCl14}, the massless supersymmetric Wess--Zumino 
model.  Even if it is far from a realistic particle physics model, the fact that we only deal with two-point functions and their simple dependence on a unique kinetic 
invariant gives a more tractable situation than more realistic theories.  Nevertheless, the presence of singularities on all integer 
points for the Borel transform of the renormalization group \(\beta\)-function is probably the generic case in massless exactly 
renormalizable QFTs, the kind we would like to better understand for their relevance in the description of our universe.

Our former studies are all based on the same simple Schwinger--Dyson equation, solved through the combination of the extraction of the 
anomalous dimension of the field from the Schwinger--Dyson equation and the use of the renormalization group equation to obtain the full 
propagator from this anomalous dimension.  We will limit ourselves to the simplest one of the Schwinger--Dyson equations, since it allows us to 
retain a degree of explicitness in the apparent explosion of different series appearing in the object named the {\em display} by Jean Ecalle, 
an object which collects all information on the alien derivatives of a function.  A proper extension of the arguments put forward 
in~\cite{BeSc12} should prove that any higher order correction to this Schwinger--Dyson equation will only change higher order terms in the 
individual components of this display, letting its main characteristic unchanged.  The factorial growth of the number of high order terms 
beyond the large \(N\), planar limit would present a further challenge, but we will see that there are plenty of questions to solve before.

The fundamental insight in~\cite{BeCl14} is that it is in the Borel plane, where the alien derivatives 
have a clear meaning as singular parts of a function at a given point, that general properties are easier to prove. 
However, most computations are easier to carry on in the form of 
transseries, where the computations look like mechanical operations on formal objects.  However, one important component in the 
computation scheme of~\cite{BeCl14} was the contour integral representation of the propagator, with its characteristic property that the possible contours 
change when considering different points in the Borel plane. It would be interesting to 
give an interpretation of these contour integrals in the formal scheme and recover how computations can be done using the 
expansion of the Mellin transform, with subtracted pole parts, at integer points.  However, we will see that the simple approximations used for this 
work do not need such a development.

We start with the renormalization group equation (RGE) for the two-point function
\begin{equation} \label{renorm_G}
 \partial_L G(a,L) = \gamma(1 +  3a\partial_a )G(a,L),
\end{equation}
where we have used $\beta=3\gamma$, which can be proved by superspace \cite{Piguet} or Hopf \cite{CoKr00} techniques.  {A derivation of this equation for the solution of a Schwinger--Dyson equation is detailed in~\cite{BeSc08}, see also~\cite{Cl15}.}
Here $L=\ln (p^2/\mu^2)$ is the kinematic parameter. Expanding $G$ in this parameter \(L\),
\begin{equation} \label{rep_G_serie}
 G(a,L) = \sum_{k=0}^{+\infty}\frac{\gamma_k(a)}{k!}L^k
\end{equation}
(with $\gamma_1:=\gamma$)\footnote{{Be aware that other authors~\cite{BrKr99,KrYe2006} use different conventions but the same $\gamma_k$ notations.}} gives a simple recursion on the $\gamma_k$s
\begin{equation} \label{renorm_gamma_old}
 \gamma_{k+1} = \gamma(1+ 3a\partial_a)\gamma_k.
\end{equation}
Therefore, at least in principle, it is enough to know $\gamma$ to rebuild the two-point function.

On the other hand, we also have the (truncated) Schwinger--Dyson equation, graphically depicted as
\begin{equation}\label{SDnlin}
\left(
\tikz \node[prop]{} child[grow=east] child[grow=west];
\right)^{-1} = 1 - a \;\;
\begin{tikzpicture}[level distance = 5mm, node distance= 10mm,baseline=(x.base)]
 \node (upnode) [style=prop]{};
 \node (downnode) [below of=upnode,style=prop]{}; 
 \draw (upnode) to[out=180,in=180]   
 	node[name=x,coordinate,midway] {} (downnode);
\draw	(x)	child[grow=west] ;
\draw (upnode) to[out=0,in=0] 
 	node[name=y,coordinate,midway] {} (downnode) ;
\draw	(y) child[grow=east]  ;
\end{tikzpicture}.
\end{equation}
{The L.H.S. is the two-point function while the R.H.S. contains two dressed propagators, which are equal to the free propagator multiplied by the two-point function.} Computing the loop integral allows to write this equation as
\begin{equation} \label{SDE_old}
 \gamma(a) = a\left.\left(1+\sum_{n=1}^{+\infty}\frac{\gamma_n}{n!}\frac{\text{d}^n}{\text{dx}^n}\right)\left(1+\sum_{m=1}^{+\infty}\frac{\gamma_m}{m!}\frac{\text{d}^m}{\text{dy}^m}\right)H(x,y)\right|_{x=y=0} =: a\mathcal{I}(H(x,y)).
\end{equation}
with $H$ known as the  Mellin transform of the one-loop integral:
\begin{equation} \label{def_H}
 H(x,y) := \frac{\Gamma(1-x-y)\Gamma(1+x)\Gamma(1+y)}{\Gamma(2+x+y)\Gamma(1-x)\Gamma(1-y)}.
\end{equation}
The idea of \cite{Be10a}, which was fully exploited in \cite{BeCl13} is to replace the one-loop Mellin transform by a truncation 
containing its singularities. Let us define 
\begin{equation} \label{form_F}
 F_k := \mathcal{I}\left(\frac1{k+x}\right)=\frac{1}{k}\biggl(1+\sum_{n=1}^{+\infty}\left(-\frac{1}{k}\right)^n\gamma_n\biggr).
\end{equation}
(which gives the contributions of the poles $1/(k+x)$ or $1/(k+y)$ of $H$) and 
\begin{equation} \label{def_Lk}
 L_k:=\mathcal{I}\left(\frac{Q_k(x,y)}{k-x-y}\right)=\sum_{n,m=0}^{+\infty}\frac{\gamma_n\gamma_m}{n!m!}\left.\frac{\text{d}^n}{\text{d}x^n}\frac{\text{d}^m}{\text{d}y^m}\frac{Q_k(x,y)}{k-x-y}\right|_{x=0,y=0}
\end{equation}
which contains the contributions from the {part of \(H\) singular on the line \(k-x-y=0\)}. Here $Q_k$ is a suitable expansion of 
the residue of $H$ at this singularity, a polynomial in the product \(xy\).

It was shown in \cite{Be10a} that these $F_k$ and $L_k$ obey renormalization group derived equations. Here we are only interested 
in the equations of $L_k$ which are
\begin{equation} \label{equa_L}
 (k-2\gamma - 3\gamma a\partial_a)L_k = Q_k(\partial_{L_1}\partial_{L_2})G(a,L_1)G(a,L_2)\Bigr|_{L_1=L_2=0} = \sum_{i=1}^kq_{k,i}\gamma_i^2.
\end{equation}
The Schwinger--Dyson equations can also be written in terms of these functions. Since we are looking for a resummation of the two-point
function along the positive real axis, and since it was shown in \cite{BeCl14} that the $L_k$ are responsible for the singularities 
of the Borel transform of $\gamma$ (and therefore of the two-point function) on the real axis, we will only take care of the terms 
of the Schwinger--Dyson equation involving $L_k$. They have the very simple form
\begin{equation} \label{SDE_phys_plan}
 \gamma(a) = a\sum_{k=1}^{+\infty}L_k(a) + (\text{contributions from }F_k){=a+O(a^2)}.
\end{equation}
In order to simplify the results of \cite{BeCl14}, we consider $\gamma$ and all the other quantities as formal series in $r:=1/(3a)$ rather than 
in $a$. We then perform a Borel transform in $r$, according to the conventions most used in the mathematical literature. The perturbative domain is then the one
for large values of \(r\) and typical nonperturbative contributions will be of the form \(e^{-nr}\) for some integer \(n\).   The advantage being that the singularities of the Borel transform $\hat\gamma$ are now located in $\Z^*$ rather than 
$\Z^*/3$. Let us notice that we now have $\hat\gamma(0)=1/3$, but what is most relevant is that \(\hat\beta(0) = 3 \hat\gamma(0) = 1\).

As explained in \cite{BeCl14}, a perturbative analysis (i.e., for $\xi$ small) of the Borel-transformed renormalization group equation suggests to
write the Borel transform $\hat G:=\mathcal{B}(G-1)$ as a loop integral
\begin{equation} \label{param_G}
 \hat{G}(\xi,L) = \oint_{\mathcal{C}_{\xi}}\frac{f(\xi,\zeta)}{\zeta}e^{\zeta L}\d\zeta
\end{equation}
where $\mathcal{C}_{\xi}$ is any contour enclosing $\xi$ and the origin. Writing the 
renormalisation group equation \eqref{renorm_G} in the Borel plane and in term of the $f(\xi,\zeta)$ function we get
\begin{equation} \label{renorm_f}
 (\zeta-\xi)f(\xi,\zeta) = \hat{\gamma}(\xi) + \int_0^{\xi}\hat{\gamma}(\xi-\eta)f(\eta,\zeta)\d\eta + \int_0^{\xi}\hat{\beta}'(\xi-\eta)\eta f(\eta,\zeta)\d\eta.
\end{equation}

\section{Resummations of the anomalous dimension}

We purposefully put a plural in the ``resummations'' of the title of this section to emphasize that two distinct resummations will be performed here.
First the median resummation and its transseries analysis deliver exponentially small terms, then we sum the dominant terms of the obtained transseries.

\subsection{Transseries solution}

We want to compute the leading coefficient of \(e^{-nr}\) in the transseries expansion of  $\beta$. Let us start by writing 
the Schwinger--Dyson equation \eqref{SDE_phys_plan} and the renormalization group-like equation \eqref{equa_L} with the variable \(r\).
We obtain, {while singling out the lowest order term coming from \(F_1\),} 
\begin{equation} \label{SDE_utile}
 \beta =\frac 1 r + \frac 1 r \sum_{k=1}^{+\infty}L_k + (\text{contributions from }F_k)
\end{equation}
and 
\begin{equation} \label{eqLk}
k {L}_k = \frac23 {\beta} {L}_k - r \beta \partial_r {L}_k
 	+ \sum_{i=1}^k q_{k,i}{\gamma}_i^2.
\end{equation}
We are interested in the freedom in the solutions of this system of equations: the perturbative solution is uniquely defined, but since it is a system of differential equations, it must have a space of solutions.  Using the fact that the two dominant terms of \(\beta\) are  \(r^{-1} - 2/3r^{-2}\), the dominant orders of the linearized equation for \(L_k\) are:
\begin{equation}
     k L_k = - \partial_r L_k + \frac 2 3 r^{-1}(L_k + \partial_r L_k) 
\end{equation}
The dominant order of the solution is:
\begin{equation}\label{dominant}
 	L_k = m_k  r^{\frac 2 3 (1-k) } e^{-k r }
\end{equation}
One can check that the additional terms coming from substituting this value of \(L_k\) in the system of equations are smaller by at least \(r^{-2}\), so that they cannot change the exponent \(\frac23(1-k)\) of this solution, but only multiply this solution by a power series in \(r^{-1}\).
Since a possible deformation of the solution proportional to \(e^{-kr}\) signals the possibility of a nonzero \(\Delta_k\), we recover the results of~\cite{BeCl14} on the possible forms of the alien derivations of \(\beta\), now written in the formal model instead of the convolutive one.

The computation of the \(r^{-1}\) corrections was carried out in~\cite{BeCl13} in the case of \(L_1\) and involves summations over the effect of all the other \(L_k\) as well as over the \(F_k\). The language was different, but the resulting computations are totally equivalent to what would be the computation of the terms proportional to \(m_1\) in a full solution.  
Such a computation nevertheless  involves summations over \(k\) which give highly nontrivial combinations of multizeta values, some of which cancel and the others can be expressed as product of zeta values. In our following work~\cite{BeCl14}, the introduction of the contour integral representation of the propagator of equation~(\ref{param_G}) gave a simple interpretation of these results and a prospective way of carrying the computations up to larger orders.
We will not consider these corrections here, since the simple study of the dominant terms gives already quite interesting results.

The nonlinear nature of the system of equations~(\ref{SDE_utile},\ref{eqLk}) means that the general solution will have terms with any product of the \(m_k\) as coefficient.  The behavior of exponentials under differentiation and multiplication ensures that such a term will also have a factor \(e^{-nr}\) with \(n\) the sum of the indices of the \(m_k\) in the coefficient.
We therefore have that \(e^{-r}\) only appears with the coefficient \(m_1\), but \(e^{-2r}\) can have the coefficients \(m_2\) or \(m_1^2\), \(e^{-3r}\), the coefficients \(m_3\), \(m_2 m_1\) or \(m_1^3\),\dots\  
The question then is to know which is the larger possible term for a given coefficient \(\prod m_k\) in the evaluation of \(\beta\). It turns out that the larger possible terms come from the product \(r \beta \partial_r L_k\) if \(L_k\) was the source of the largest term in \(\beta\) with the same coefficient. 
The end result is that the largest power of \(r\) coming in \(L_k\) for some product of \(m_j\) including \(m_k\) is \(\frac2 3 \sum (1-j)\), giving a term with an exponent less for \(\beta\).  The nice point is that the dominant term among the ones with the factor \(e^{-nr}\) is the one proportional to \(m_1^n\), which has no additional powers of \(r\) for \(L_1\) and just the factor \(r^{-1}\) for \(\beta\).

We therefore can parameterize the sum of the dominant terms in \(L_1\) with
\begin{equation}\label{l1trans}
	L_1 = \sum_{n=1}^\infty c_n m_1^n e^{-nr}
\end{equation}
Using that at this order, \(\beta\) is \(r^{-1} + r^{-1} L_1\), the equation~(\ref{eqLk}) for \(k=1\) gives the following recurrence relation for the \(c_n\):
\begin{equation}\label{relation_c}
	(1-n) c_n = \sum_{p=1}^{n-1} c_{n-p}\, p\, c_p
\end{equation}
If we define a formal series \(S\) by
\begin{equation} \label{series_S}
 S(x) := \sum_{n\geq1}c_nx^n.
\end{equation}
we obtain that a first transseries solution for \(L_1\) is given by
\begin{equation}
	L_1 = S(m_1 e^{-r})
\end{equation}

\subsection{Summation of the transseries}

The previous subsection introduced the formal series \(S\), and we need to know its properties, in particular its radius of convergence.
The inductive formula \eqref{relation_c} for the $c_n$'s implies a differential equation for $S(x)$ (seen as a formal series):
\begin{equation*}
	\frac {S(x)} {x} - S'(x) = S(x)S'(x).
\end{equation*}
Dividing by \(S(x)\) and regrouping terms depending on \(S\), one obtains:
\begin{equation*}
	\frac{S'(x)}{S(x)}  + S'(x) = \frac{1}{x}.
\end{equation*}
The left hand side is the logarithmic derivative of the function $F(x):=S(x)e^{S(x)}$ so that we have
\begin{equation*}
	S(x) e^{S(x)} = k x,
\end{equation*}
for some $k\in\mathbb{R}$. From the definition of \(m_1\) in~\eqref{dominant} and the comparison with the formula~\eqref{l1trans}, we see that \(c_1=1\), which also fixes \(k=1\). The presence of a minimum of the function \(u \rightarrow u e^u\) for \(u=-1\) with the value \(-1/e\) gives rise to 
a singularity of \(S(x)\) for \(x = -\frac 1 {e}\) of the square root type. Since it is the singularity nearest to the origin, it implies that the convergence radius of the series is \(\frac1 {e}\).

In fact, the above function inversion problem has been studied and the solution of the case \(k=1\) is known as (the principal branch of) Lambert's \(W\)-function.
Using the initial condition $S'(0)=1$ and the fact that $W'(0)=1$ we find that the series $S(x)$ of \eqref{series_S} is actually
\begin{equation}
 S(x)=W(x).
\end{equation}
An explicit series representation of the Lambert \(W\)-function is known :
\begin{equation*}
 W(x) = \sum_{n\geq1}\frac{(-n)^{n-1}}{n!}x^n.
\end{equation*}
This formula can be deduced from the Lagrange inversion formula. The convergence radius $1/e$ is then a simple consequence of the Stirling formula for the factorial.

All in all we have shown that the {sum of the lowest order terms in all nonperturbative sectors of the anomalous dimension} is
\begin{equation} \label{result_anormal_dimension}
 \gamma^\mathrm{res}(r) = r^{-1} W( m_1 e^{-r}) + \mathcal{O}(r^{-2})
\end{equation}
and is defined in the region $|m_1 e^{-r}|<1/e$ of the complex plane. 

{The construction presented here is a particular example of ``transasymptotic analysis'', which suggests that similar formulae exist at any order in $e^{-r}$. See for example \cite{Costin2001} and references therein.} {An interesting aspect of the analysis in~\cite{Costin2001} is that they show that the singularity of this lowest order resummed solution signals a singularity of the full solution in its vicinity.}

\subsection{Links with the alien calculus}

In the preceding sections, we studied a possible transseries deformation of the perturbative solution for the \(\beta\)-function, but to what use can it be put for the evaluation of the function? In particular, could it be possible to have a determination of \(m_1\)?  
Part of the response comes from the idea of the bridge equations.  Since the Stokes automorphism and its powers respect products and commutes with the derivation with respect to \(r\), the functions remain solution of the equations when transformed by these automorphisms.  Since in the formal model, the alien derivation \(\Delta_n\) gives rise to a factor \(e^{-nr}\),  the solutions after the action of a Stokes automorphism will be in the form of a transseries.

Since the most general transseries solution is a function of the parameters \(m_k\) appearing in the linear deformations of the solution~\eqref{dominant}, 
alien derivatives can be expressed through derivations acting on these parameters, giving bridges between alien calculus and ordinary calculus.  The alien derivation \(\Delta_n\) is, a priori, any combination of operations which lower the weights by \(n\),  so that for example, \(\Delta_1\) has not only a term proportional to \(\partial/\partial m_1\), but also \(m_1 \partial/\partial m_2\) and many others.  Nevertheless, the same reasons which made the terms proportional  to \(m_1^n\) dominate imply that the coefficient \(f\) of \(\partial/\partial m_1\) is the most important part of \(\Delta_1\).  A value for this coefficient \(f\) could be extracted in~\cite{BeCl13} from the comparison of the asymptotic behavior of the perturbative series for \(\gamma\) deduced from the singularities of the Borel transform and the known coefficients of this series. 

The dominant term in the singularity at the point~\(n\) necessarily comes from the coefficient of \(m_1^n\) and can only be extracted by the \(f^n (\partial/\partial m_1)^n\) term in \(\Delta_1^n\).  The \(n!\) coming from the iterated differentiations is compensated by the \(1/n!\) factor in front of \(\Delta_1^n\) in the definition of \(\Delta^+_n\), so that we obtain, from the relation between alien derivatives and singularities of the Borel transform, that \(\hat L_1\) has a pole in \(n\) with residue \(c_n f^n\), while the only divergent part for \(\hat \beta\) is proportional to \(- c_n f^n \log(|\xi - n|) \).

These singularities of the Borel transform are transmitted to the median average. In turn, these singularities of the integrand produce nonperturbative contributions to the result of resummation, so that the factors like \(f\), determined through alien calculus, can be used to fix the unknown coefficients in the transseries expansion.  What is important is that the consistency of all the steps of the resummation procedure with the products and derivation ensures that the result respects the original equations and must therefore be of a form compatible with the transseries solution.  The influence of the singularity at 1 will therefore be sufficient to obtain the dominant part for the singularities for all \(n\). 
 
Although we have been able to compute nonperturbative terms, with coefficients which could be computed from the perturbative expansion of the anomalous dimension, the situation seems quite complicated. 
Indeed the resurgent analysis seems to make the situation go from bad to worst: instead of a unique formal series, we end up with formal series multiplying \(e^{-rn}\) for each \(n\), and with furthermore coefficients which are polynomials in \(r^{-2/3}\)  and \(\log r\) of degrees growing with \(n\), with many undetermined coefficients. 
Moreover, each of these series are actually divergent and need some form of resummation.
However, we may remember that in many cases, divergent series are not so bad news, and as Poincar\'e has put it, they are ``convergent in the sense of astronomers'': the first few terms give a fairly accurate approximation of the final result, as is the case for example in quantum electrodynamics.
Our position will therefore be to use the information we have and forget for the time being about all the unknown quantities. We would of course prefer to have arguments proving that indeed what we neglect is negligible, but it is the best we can do at the moment.
  
{We reshuffle the transseries and write them as series in $r$ whose terms are series in $e^{-r}$. This operation is inspired by a remark of Stingl \cite{St02}, page 70 about 
physical} considerations on what the ``true'' observables {are,} were put forward to provide a justification to this manipulation.
The take home message could be that it is important to keep all the terms of a convergent series but series with 0 convergence radius could be truncated without remorse.  A physicist way of dealing with such a situation would be to look at how the results change when we add terms from the formal series, but we would need at least one more term.

A more mathematical view could come from transmonomials~\cite{Sa07}, which are special functions with simple properties under the action of alien derivatives.
This could lead to an expansion of the function where transmonomials get multiplied by ``alien constants'', functions on which all alien derivatives give zero and therefore easily computable from their power series expansion.  The simplest transmonomial \(\mathcal U^1\), with the only non zero alien derivative \(\Delta_1 {\mathcal U}^1 =1 \) would replace \(e^{-r}\) in all our transseries expressions and take care of the dominant terms at large orders neglected in the naive approach.

To conclude this short exposition of the idea of resumming the transseries, let us emphasize that in other contexts, methods using grouping of terms of nonsummable families have been put to good use to produce convergent expressions.  
Arbitrary groupings can produce arbitrary results, but some well defined procedures have been shown to reproduce the results of a Borel summation.
A prime example is the arborification procedure presented in Ecalle's work on mould calculus~\cite{Ecalle1992}, which separates terms in smaller parts to be regrouped in other objects. {Our procedure might be seen as an 
arborification where only ladder trees give nonzero contributions. The reader can be referred to \cite{FaMe12}, Section 6} for a clear introduction of the arborification--coarborification 
in the context of linearization problems. Other cases appear in the study of Dulac's problem~\cite{Ec92}. The main advantage of such procedures is that they allow practical computation. It is thus possible that the somewhat ad hoc computation presented above can also be justified.

\section{Properties of the Green function}

\subsection{Nonperturbative mass scale generation}

The mass of a particle is given by the position of a pole of the two-point function as a function of the invariant \(p^2\) of the momentum.  In our case, there is always a pole for \(p^2=0\), reflecting the fact that we started from a massless theory, since the function \(G\) we are studying is a multiplication factor for the free propagator, the same for all states of the supermultiplet.  We define a mass scale as the value of the external momentum $p$ for which the Green function has a singularity.  The expansion of \(G\) in powers of the logarithm \(L= \log(p^2/\mu^2)\) is ill suited for such an analysis. We expect the sign of \(p^2\) to fundamentally change the situation, the pole for a physical particle being for a timelike \(p\), while \(L\) only change by \(i\pi\) when going from timelike to spacelike momenta.

This is why we will rather use the integral representation \eqref{param_G}:
\begin{equation*}
 \hat{G}(\xi,L)=\oint_{\mathcal{C}_{\xi}}\frac{f(\xi,\zeta)}{\zeta}e^{\zeta L}\d\zeta.
\end{equation*}
It was shown in \cite{BeCl14} that the function $\zeta\mapsto f(\xi,\zeta)$ has singularities at $\zeta=\xi$ and at \(\zeta=0\). Therefore we can expand the contour 
$\mathcal{C}_{\xi}$ to infinity without changing the value of the integral. This being done, we can make the lateral alien derivative go 
through the integral and obtain:
\begin{equation*}
 \Delta_n^+\hat{G}(\xi,L)=\oint_{\mathcal{C}}\frac{\Delta_n^+f(\xi,\zeta)}{\zeta}e^{\zeta L}\d\zeta,
\end{equation*}
where the lateral alien derivative acts on the $\xi$ variable. Taking the lateral alien derivative of the renormalisation group equation 
\eqref{renorm_f} we get
\begin{equation} \label{renorm_f_lat_alien_der}
 (\zeta-(\xi+n))\Delta_n^+f(\xi,\zeta) = \frac 1 3 \Delta_n^+\hat\beta(\xi) + \frac 1 3 \sum_{i=0}^n \bigl(\Delta_{n-i}^+\hat{\beta}\star\Delta_i^+ f \bigr)(\xi,\zeta) + \sum_{i=0}^n \bigl(\Delta_{n-i}^+\hat{\beta}'\star \text{Id}. \Delta_i^+f\bigr)(\xi,\zeta).
\end{equation}
We are looking for the dominant term in $\Delta_n^+f(\xi,\zeta)$ seen as a function of $\xi$. It will come from the dominant term in \(\Delta_n^+ \hat\beta\), which is constant in \(\xi\), while all other terms give higher powers of \(\xi\). It will therefore be given by a term \(f_{n,0}(\zeta)\), which satisfies
\begin{equation*}
 (\zeta-n)f_{n,0}(\zeta) = c_n f^n.
\end{equation*}
Plugging this into the integral representation of $\Delta_n^+\hat{G}(\xi,L)$ we get
\begin{equation*}
 \Delta_n^+\hat{G}(\xi,L) = \frac{c_n f^n}{n}\left(e^{nL}-1\right)+\mathcal{O}(\xi^{2/3}).
\end{equation*}
The transseries expansion of \(G\) deduced from these singularities of \(\hat G\) is then 
\begin{equation*} 
 G^\mathrm{res}(r,L)= 1+\frac 1 r \sum_{n=1}^{+\infty}\frac {c_n} {n} (f e^{-r})^n \Bigl(e^{nL} -1 \Bigr) + \text{higher orders}
\end{equation*}
Now, as we have done 
for the anomalous dimension in the previous section we can simply sum the above series, without worrying on the other terms. 
If we neglect 1 with respect to \(e^{nL} = (p^2/\mu^2)^n\), we obtain a function of \(f e^{L-r}\) with Taylor coefficients \(c_n/n\).
Since the \(c_n\) are the Taylor coefficients of Lambert's \(W\)-function, these are the coefficients of the primitive of the \(W\)-function divided by \(x\).
This is a function which grows only logarithmically for positive arguments, but goes to 0 has a three half power of the variable at \(-1/e\).
$f$ was numerically computed in \cite{BeCl13} to be $0.208143(4)$. This shows that for an euclidean momentum where \(e^L\) is positive, we are in a situation where the function grows really slowly and the large terms of the series, proportional to \( (p^2)^n \),  combine to a simple logarithmic correction to the propagator. 
Since propagators can be Wick rotated to the euclidean domain in loop computations, this means that nothing prevents us, at this approximation level, from defining consistently the renormalized theory.  
On the other hand, we have for a finite value of \(p^2\) in the timelike domain a singularity of the propagator which defines a mass scale
\begin{equation}
 M_{NP}(r)^2 = \frac{\mu^2}{f}e^{r-1}.
\end{equation}
Let us notice that we find that the nonperturbative mass goes to infinity as $r$ goes to infinity, which corresponds to $a$ going to $0^+$. 
This was to be expected. However this mass scale is not renormalization group invariant, since a renormalization group invariant mass scale should involve a factor \(r^{2/3}\). We hope that a more careful analysis can give back this factor so that we obtain a fully consistent analysis of this nonperturbative mass scale.

Let us finally remark that the detour by the contour integral representation of \(G\), which is very valuable if we wanted to consider all the corrections proportional to powers of \(L\) in the expansion of \(G\), is not really necessary at this level of approximation. We could have simply deduced that a term proportional to \((p^2)^n\) appears when considering a \(e^{-nr}\) term in the \(\beta\)-function.  

\subsection{Analyticity domain: a necessary acceleration?}

We want to see how the resummed two-points function $G^{\text{res}}$ could be obtained through Laplace transforms, to better understand its analyticity domain. 
We make the bold approximation that the size of the Borel transform at a point can be approximated by the contribution of the nearest singularity. We have seen that the singularity in the point \(n\) is dominated by the contribution \(1/n! \Delta_1^n \hat G\) in the lateral derivative, which has the factor \(c_n (p^2)^n\). 
The coefficients \(c_n\) have also a power like behavior so that the domain in \(r\) where the Laplace transform of \(\hat G\) is well defined shrinks when \(p^2\) grows.
Since we would like that our theory defines the two-point function for any values of the momentum, we cannot define it by a simple Laplace transform. {Indeed, even if the reformulation of the Schwinger--Dyson equation we use does not make it explicit, the proper definition of the two-point function is necessary to compute the loop integral appearing in the definition of the \(\beta\) function. It is therefore important to have at all stages computation which are uniform in \(p\).}

The fact that the dominant terms in the transseries representation sum up to an analytic function of \(p^2\), with a well behaved extension to any positive values of \(p^2\) is of no relevance here: the Laplace transform is well defined only if \(p^2\) is small enough that we are in the convergence domain of the sum of the dominant terms. 
We need
 \begin{equation*}
 |e^{-r}|\leq e^{\kappa}=|c|^{-1} e^{-L+1},
\end{equation*}
in other terms that the real part of \(r\) should be larger than \(-\kappa\), which grows like \(L\).

In order to be able to do the Laplace integral, we could think of doing a Borel transform with respect to \(p^2\), but we cannot see how it could be possible to use the Borel-transformed two-point function as an ingredient of the Schwinger--Dyson equations.  A possible way out is rather through acceleration. This formally corresponds to making a Laplace transform followed by the Borel transform of the function expressed in terms of a new variable, which is a growing function of the old one.  If the Laplace transform was well defined, one sees that there exists a kernel \(K(\xi_1,\xi)\) such that the new Borel transform \(\hat f_1\) of a function \(f\) is given by:
\begin{equation}
	\hat f_1(\xi_1) = \int_0^\infty K(\xi_1,\xi) \hat f(\xi) d\xi.
\end{equation}
For fixed \(\xi_1\), the kernel \(K(\xi_1,\xi)\) vanishes faster than any exponential when \(\xi\) goes to infinity, allowing this acceleration transform to remain defined in cases where the Laplace transform in the original variable was not possible.
In fact, since the Borel transform remains of  exponential growth, but only with a coefficient which can be arbitrarily large, the acceleration transform can be defined whenever its kernel has a slightly faster decay than the exponential. This is the case already for the kernel associated with the change of variable \(r \to r_1 = e^r\), for which the kernel behaves like \(\exp(-\xi \log\xi)\) for large \(\xi\) and fixed \(\xi_1\).

The terms we kept of the transseries expression of the  two-point function translate in the following value for its accelerated Borel transform in terms of the Borel transform \(\hat W\) of the Lambert function:
\begin{equation}
	 \hat G_1(\xi_1, p^2) \simeq \hat W ( \xi_1 p^2 )
\end{equation}
Since the Lambert function is holomorphic in a neighborhood of the origin, its Borel transform is an entire function, so that \(\hat G_1\) is well defined {at this approximation level}.  However, if the final Laplace transform is to be valid for any value of \(p^2\), it must be done in directions such that \(\hat W\) is smaller than any exponential, and this in turn restrict the angular width of the domain in \(r_1=e^r\) where the field theory can be defined by resummation.
In turn, this implies that the imaginary part of \(r\) is bounded: the limits of the analyticity domain are lines of fixed imaginary part, which correspond to circle arcs tangent to the  real axis in the original coupling \(a\).
The ensuing analyticity domain is quite similar to the one proposed by 't Hooft~\cite{Ho79} for any sensible quantum field theory.

It remains to know whether for some functional of the two-point function singularities of the Borel transform in the variable \(\xi_1\) could appear, ensuring that there are non trivial alien derivatives in this second Borel plane {and thus proving the unavoidable character of acceleration.  We must remember that the approximation of the system of equations obeyed by the renormalization group function used in this work is a rather crude one. Conceptually, we do not have a clearly defined differential system, but a functional equation involving the two-point function with its dependence on the momentum.  For the perturbative solution, we could transform it to an infinite system of differential equations by considering the function to be given by its Taylor series at the origin in the variable \(L\), but this approximation is not suitable for the computation of the properties of the analytic continuation of the Borel transform.  We had to introduce terms associated with the poles of the Mellin transform to obtain deformation parameters naturally associated with each of the terms \(e^{nr}\), with \(n\) any nonzero integer. But our system of differential equations has other components, since we also need the coefficients \(\gamma_k\). This suggests that the terms we discussed in~\cite{BeCl14} do not exhaust the possible transseries deformations of the solution, but the change in the representation of the two-point function which would allow to pinpoint these additional possible terms goes far beyond the ambition of this work.}

{In fact, this acceleration procedure should not be viewed with too much fear.   It can be seen as a tool to transform the difficult problem of bounding the analytic continuation of the Borel transform into the much simpler problem of finding some kind of formal solutions which allow to characterize the possible alien derivatives in a second Borel plane.  In any case, what happens in the different Borel planes are somehow unrelated, so that the process of analytic continuation, analysis of the singularities and eventually averaging of the Borel transform is totally independent of the fact that it will be followed by a Laplace transform or an acceleration transform.}

\section*{Conclusion}

Using to a bigger extent the power of alien calculus and transseries expansions we have been able to go much further than in our previous work~\cite{BeCl14}. The dominant terms of the singularities of the Borel-transformed anomalous dimension of the theory has been computed.
This computation was carried out by using transseries expansions and considering the effect 
of lateral alien derivatives on the Schwinger--Dyson and the renormalization group equation of the theory. Then, using a suitable form 
of the median resummation, this gave the first order in every `instantonic' sectors of the anomalous dimension of our theory.

Following a procedure suggested by Stingl, we have kept in this transseries solution the terms of a formal series in 
$e^{-r}$, forgetting the terms in (negative) powers of \(r\). This series of the dominant terms turned out to be convergent and its sum evaluated.

The same analysis could also be transferred to the two-point function of the theory. Indeed, the two-point function is 
essentially determined by the anomalous dimension through the renormalization group. 
The important point is that the singularity of the anomalous dimension in the point \(n\) of the Borel plane, or a term \(e^{-nr}\) in the transseries expansion, gives rise to a term proportional to \((p^2)^n\) for the two-point function.  What looked like negligible contributions to the anomalous dimension becomes dominant in the two-point function for large \(p^2\). 
While an explicit value of the factor multiplying \(p^2\) could not be 
obtained, the functional dependence can be obtained.  In the euclidean domain, which corresponds to \(p^2\) positive with our conventions, the asymptotic behavior of the Lambert function means that {this series of powers of 
\(p^2\) has a finite radius of convergence.} These powers of \(p^2\) sum up to something of merely logarithmic growth. 
This final asymptotic behavior may however appear only for so large values of the momentum that it would be totally invisible in usual nonperturbative studies, where the ratio between the largest and the smallest scales that can be studied is rather limited.
Nevertheless the two-point function has, at this approximation level, a quite regular behavior at the smallest scales, in contradistinction to the divergence for some finite scale obtained with finite order approximations of the \(\beta\)-function: many general arguments for the triviality of such a quantum field theory
with a positive \(\beta\)-function break down.  We must however remain cautious, since we are just scratching the surface of the kind of beyond perturbative theory analysis resummation theories allow, and many new phenomena may be revealed by a more careful study.

The simple analytic dependences on \(p^2\) of all the terms in the expansion of the two-point function make it easy to study its analytic continuation to the timelike domain, with negative \(p^2\).  In this case, a singularity appears which is of square root type.  This singularity defines a nonperturbative mass scale for the theory, but our computation is not fully satisfying since this mass scale is not fully renormalization group invariant. 

The fact that even our fairly simple computation revealed a nonperturbative mass scale for the theory is quite remarkable. 
Also we did not have to choose the functional dependence of the two-point function, but it was provided by our computation. In our case, singularities of the Borel transform were associated with ultraviolet divergent contribution to the two-point function: 
in an asymptotically free theory like QCD, we would obtain infrared divergent contributions, but likewise it could be possible to sum this contributions to obtain well behaved propagators up to the lowest scales. Since we do not decide a priori their functional form, we may have signs of confinement in the form of states having singularities different than poles in the timelike domain. 

Considerations about the way that the sum of the terms of the transseries could come from the Laplace integral give indications on the growth 
of the Borel transform, which is harder to check: changing the 
variable on which the final Laplace transform is done corresponds to an integral transform (dubbed acceleration by Ecalle) with a kernel 
that have faster than exponential decay at infinity. A suitable change of variables therefore allows to define a new germ of analytic function 
near the origin, the singularities of which can be studied by a new set of alien derivations. If all such possible accelerations do not present 
singularities, the first Borel transform should be suitable to directly define the sum, giving an indirect check on the growth of the Borel transform. 

Limits on the analyticity domain for complex values of the 
coupling constant devised by 't Hooft~\cite{Ho79} and Stingl~\cite{St02} suggests, contrarily to what Stingl said in some papers, that at least 
one such acceleration should be needed in the case of nonabelian gauge theories. 
This question is linked to the shape of the analyticity domain in the coupling constant of the theory, that we did not study.  However, in the case of the two-point function, if \(r\) is given an imaginary part of \(i\pi\), \(e^{-r}\) changes sign and the singularity which was for timelike momenta enters the euclidean domain. 
This should pose serious problems to the continuation of the theory to such values of the coupling, so that \(r\) should be limited to a band of finite extension in the imaginary direction, which converts to the horn-shaped domains proposed by
't Hooft when converting back to the coupling proportional to \(r^{-1}\). 

It should certainly be interesting to compute the higher order corrections to the transseries solutions of the system of equations studied here and try to deduce their full system of alien derivations. 
An effective computation however appears to be quite a formidable task, but in what are certainly simpler cases, mould calculus has been shown to provide for quite explicit results, expressing results in terms of resurgent monomials~\cite{Sa07}.
With this strategy, one could avoid as much as possible to work explicitly in the Borel plane, even if the analytic continuation of functions in the Borel planes (plural if acceleration is needed) is the ultimate justification of the computations one may attempt.

Moreover, our study could also be carried out when including additional terms involving higher-loops primitively divergent diagrams 
in the Schwinger--Dyson equation. This should allow to expand the results of \cite{BeSc12}, which only considered the asymptotic behavior of the perturbative series, that is, the singularities closest to the origin of the Borel transform.

In this work, we limited ourselves to the two-point functions, which have a simple dependence on a unique Lorentz invariant: a general study of quantum field theory would certainly benefit from a careful investigation of the analytic properties of the Borel-transformed Green functions in all their variables.

Finally, let us notice that the probable usefulness of mould calculus in the realm of quantum field theory expands the 
list of elements of Ecalle's theory of resurgence which should be used in physics: we have used alien calculus, median resummation 
and it seems very likely that acceleration will be needed in the next steps of our program. Bridge equations are nowadays a common tool of some 
physicists (see e.g., \cite{AnScVo12}) and we argued that we might also use mould calculus while resurgent monomials will come into the game. Proper use of these tools could well provide solutions to old questions in quantum field theory.
\bibliographystyle{unsrturl}
\bibliography{renorm}

\end{document}